%% file: main.tex
\documentclass{sig-alternate-05-2015}
\usepackage{etex}
\usepackage[
bookmarks=false
,bookmarksnumbered=true
,hypertexnames=false
,breaklinks=true
]{hyperref}
\hypersetup{
  pdfauthor={Dehghani, Azarbonyad, Kamps, Marx},
  pdftitle={Generalized Group Profiling for Content Customization},
  pdfsubject={CHIIR'16 submission},
  pdfkeywords={H.3 [Information Storage and Retrieval]},
  pdfcreator={LaTeX with hyperref package},
  pdfproducer={pdflatex}
}
\usepackage{algpseudocode}
\usepackage{algorithm}
\usepackage[table]{xcolor}
\usepackage{graphicx}
\usepackage{subcaption}
\usepackage{tikz-qtree}
\usepackage{multirow}
\usepackage{multicol}
\usepackage{amsmath}
\usepackage{mathabx}
\usepackage{MnSymbol}
\usepackage{pgfplots}
\usepackage[utf8]{inputenc}
\usepackage{amsmath}
\usepackage{MnSymbol}
\usepackage{microtype}
\usepackage{nicefrac}
\usepackage{array}
\usetikzlibrary{patterns}
\pgfplotsset{compat=newest}
\usepackage[font=small,labelfont=bf]{caption}
\newcommand{\todo}[1]{\textcolor{red}{#1}}

\makeatletter

\DeclareCaptionLabelFormat{numberless}{\ALG@name#1}
\makeatother

\def\:{\hskip0pt} 
\newcommand{\mypar}[1]{\medskip\noindent\textbf{#1}~}
\newcounter{todocnt}
\renewcommand{\todo}[1]{%
{\refstepcounter{todocnt}%
\textcolor{blue}{$\bullet\bullet\bullet$\ \sf To do (\thetodocnt): #1}}}
\renewcommand{\todo}[1]{}
\newcounter{latercnt}

\usepackage[square,comma,numbers,sort&compress,sectionbib]{natbib}
\usepackage{booktabs} 
\usepackage{tabularx} 

\def\sharedaffiliation{
\end{tabular}
\begin{tabular}{c}}

\begin{document}
\CopyrightYear{2016}
\setcopyright{acmlicensed}
\conferenceinfo{CHIIR'16,}{March 13--17, 2016, Carrboro, NC, USA}
\isbn{978-1-4503-3751-9/16/03}\acmPrice{\$15.00}
\doi{http://dx.doi.org/10.1145/2854946.2855003}
\clubpenalty=10000 
\widowpenalty = 10000

\title{Generalized Group Profiling for Content Customization}
%
%
%
%
%

\numberofauthors{1} 
%
\author{
%
%
\alignauthor
Mostafa Dehghani\(^1\) 
$\qquad$
Hosein Azarbonyad\(^2\)
$\qquad$
Jaap Kamps\(^1\)
$\qquad$
Maarten Marx\(^2\)\\
\sharedaffiliation
\affaddr{\mbox{}\(^1\)Institute for Logic, Language and Computation, University of Amsterdam, The Netherlands}\\
\affaddr{\mbox{}\(^2\)Informatics Institute, University of Amsterdam, The Netherlands}\\
        \small \{\url{dehghani,h.azarbonyad,kamps,maartenmarx}\}\url{@uva.nl}
}

\newcommand{\maingoal}{to develop a language model for a group of users based on the group preferences, which contains all, and only, essential shared commonalities of the group members, and to employ such a group profile to customize content suggestion for individual users}
\newcommand{\rqone}{How to estimate models for group profiles capturing exactly the shared commonalities of their members?}
\newcommand{\rqtwo}{How effective are group profiles to customize content suggestion for individual users?}
\newcommand{\rqthree}{How does the user's group granularity affect the quality of the group's profile?}

\maketitle
\begin{abstract}

There is an ongoing debate on personalization, adapting results to the unique user exploiting a user's personal history, versus customization, adapting results to a group profile sharing one or more characteristics with the user at hand.
Personal profiles are often sparse, due to cold start problems and the fact that users typically search for new items or information, necessitating to back-off to customization, but group profiles often suffer from accidental features brought in by the unique individual contributing to the group.
In this paper we propose a generalized group profiling approach that teases apart the exact contribution of the individual user level and the `abstract' group level by extracting a latent model that captures all, and only, the essential features of the whole group.

Our main findings are the followings.
First, we propose an efficient way of group profiling which implicitly eliminates the general and specific features from users' models in a group and takes out the abstract model representing the whole group. 
Second, we employ the resulting models in the task of contextual suggestion. We analyse different grouping criteria and we find that group-based suggestions improve the customization. 
Third, we see that the granularity of groups affects the quality of group profiling. We observe that grouping approach should compromise between the level of customization and groups' size.

\end{abstract}

%
%
%

\medskip
\small


\vspace{1mm}
\noindent
{\bf Keywords:} 
Group Profiling, Contextual Suggestion, Content Customization.  
\normalsize

\section{Introduction}

Context is pervasive on the modern web, due to cloud-based and mobile applications, making every information access interaction part of an eternal user session.  Effective ways to leverage this context are key to further enhancing the user experience, both in terms of better quality of results as in terms of easier ways to articulate complex information needs.  This requires both effective ways of personalization to an individual user as well as customization to a profile based on groups of users.
For group level analysis, there is a need for extracting a group profile that captures the essence of the group, separate from the sum of the profiles of its individual members. This profile should be ``specific'' enough to distinguish the preferences of the group from other groups, and in the same time, ``general'' enough to capture all shared tastes, expectations, and similarities of its members. 

Group profiling can help understand both explicit groups, like Facebook groups, and implicit groups, like groups extracted by community detection algorithms. There is a wide range of applications for group profiling, like understanding social structures~\citep{Tang:2011}, network visualization, recommender systems~\citep{Hu:2014,Shang:2014,Amer-Yahia}, and direct marketing~\citep{Custers:2003}. 
One of the important applications of group profiling is in the content customization problem.  Content customization is the process of tailoring content to individual users' characteristics or preferences. However, using individual preferences for content customization is not always possible. For example sometimes there is a new user in the system with no historical interactions and no rich information about the preferences, or sometimes the user is not able to determine his/her preferences explicitly. In these situations, group based content customization would be beneficial to suggest content to the user based on the preferences of the groups that the user belongs to. 

The main aim of this paper is \textsl{\maingoal}.
We break this down into three concrete research questions:
\begin{enumerate}
  \setlength{\itemsep}{0pt}
  \setlength{\parskip}{0pt}
  \setlength{\parsep}{0pt}
\item[\textbf{RQ1}] \textsl{\rqone}
\item[\textbf{RQ2}] \textsl{\rqtwo}
\item[\textbf{RQ3}] \textsl{\rqthree}
\end{enumerate}

There is various research done on the task of group profiling which, given the individual attributes and preferences, aims to find out group-level shared preferences~\cite{Senot:2011,Masthoff:2011}.
\citet{Tang:2011} presented three different methods for group profiling: \textsl{Aggregation}, which tries to find features that are shared by the whole group; \textsl{Differentiation}, which tries to extract features that can help to differentiate one group from others; and \textsl{Egocentric differentiation}, which tries to extract features that can help to differentiate members of one group from the neighbour members. In recent work, \citet{Hu:2014} proposed a deep-architecture model to learn a high level representation of group preferences.

For group recommendation there is research on building a model of a group by forming a linear combination of the individual models~\citep{Jameson:2007}. Some of them construct the group's preference model on the basis of individual preference models, using a notion of distance between preference models~\citep{Yu:2006}. Some approaches try to divide the group into several categories of homogeneous users and specify the preference model for each subgroup. Then they create the group model as a weighted average of the subgroup models, with the weights reflecting the importance of the subgroups~\citep{Ardissono:2003}.

In this paper, we assume that the model of a user's preferences is a mixture of general, specific and his/her group preferences, and we try to extract the latent group model. 
We utilize the extracted profile in the task of contextual suggestion to improve content customization based on a user's group memberships.

The rest of the paper is structured as follows.
First, in Section~\ref{sec:gp}, we explain our approach for group profiling in detail. Section~\ref{sec:exp} presents the results of our experiments on the task of group-based contextual suggestion as well as some analysis on the effect of group granularity on the quality of group profiling. Finally, Section~\ref{sec:con} concludes the paper and discusses possible extensions and future work.

\section{Group Profiling}
\label{sec:gp}

In this section, we investigate our first research question: ``\rqone''\ 

Group profiling refers to the task of extracting a descriptive model for a group of users which addresses the particular aspects that bind group members together. 
The goal of the proposed method for group profiling is to extract the latent common language model of the group which represents the shared group preferences. 
We assume that there are three different models from which the group members sample to express their preferences: \emph{group} model, \emph{general} model, and \emph{specific} model. The group model is supposed to represent the shared preferences of the group. The general model represents the general terms that anybody may use (very common observed terms), and specific model represents the terms that are attached to an individual user's preferences but not others (partially observed terms).

Each model is a distribution over terms. We consider collection language model $\theta_c$ as the general model: 
\begin{equation}
p(t|\theta_c) = \frac{c(t,C)}{\sum_{t' \in V} c(t',C)}
\end{equation}
This way, terms that are well explained in the collection model get high probability and are considered as general terms. To estimate the probability of terms given the specific model, $\theta_s$, we use the Equation~\ref{eq:Spec} and normalize all the probabilities to form a distribution: 

\begin{equation}
P(t|\theta_s) = \sum_{u_i\in G} \bigg(P(t|\theta_{u_i}) \prod_{\substack{u_j\in G \\ j \neq i}} (1-P(t|\theta_{u_j}))\bigg) \cdot idf_G(t)
\label{eq:Spec}
\end{equation}

where ${u_i}$ is the user $i$ in group $G$, and $\theta_{u_i}$ is the language model representing the user's preferences. Here, $idf_g(t)$ represents inverse document frequency of term $t$ in group $G$. 
Equation~\ref{eq:Spec} calculates probability of term $t$ to be an specific term. To this end, it simply considers the probability of a term to be important in one of the user models but not others, marginalized over all user models as well as considering group document frequency. In this way, terms that are well explained in only one user model but not others get high probability and are considered as specific terms.

Based on the generative model, each term in a user model is generated by sampling from a mixture of these three models\:---\:group, general (or collection),  specific\:---\:independently. Thus, the probability of generating term $t$ in user model $u$ would be:
\begin{equation}
p(t|u) = \lambda_{u,g} p(t|\theta_g) + \lambda_{u,c} p(t|\theta_c) + \lambda_{u,s} p(t|\theta_s),
\end{equation}
where $\lambda_{u,x}$ stands for $p(\theta_x|u)$ which is the probability of choosing model $\theta_x$ given the user $u$. 

The goal is to fit the log-likelihood model of generating all terms in the user models to discover the exact term distribution of the group model, $\theta_g$. Let $G = \{u_1, \ldots u_n\}$ be a group of users. The log-likelihood of the group would be:
\begin{equation}
\log p(G|\Lambda) = \sum_{u \in G}\sum_{t \in V} c(t,u) \log \big(\sum_{x\in\{g,c,s\}}\lambda_{u,x} p(t|\theta_x)\big),
\end{equation}
where $c(t,u)$ is the frequency of term $t$ in user model $u$ and $\Lambda$ determines the set of all parameters:
\begin{equation}
\Lambda = \{\lambda_{u,c}, \lambda_{u,s}, \lambda_{u,g} \}_{u \in G} \cup \{\theta_g\}
\end{equation}

As we have mentioned, we estimate $\theta_c$  and $\theta_s$ based on the collection language model as well as the patterns of terms occurrences in the documents and we make them fixed in the model. Finally, to fit our model, we estimate the parameters using the maximum likelihood (ML) estimator. So we solve the following problem:
\newcommand{\argmax}{\arg\!\max}
\begin{equation}
\Lambda^* = \argmax_\Lambda p(G|\Lambda)
\end{equation}

Assuming $X_{u,t}\in \{g,c,s\}$ as a hidden variable indicating which model has been used to generate term $t$ in user model $u$, we can compute the parameters efficiently using Expectation-Maximization (EM) algorithm. The stages of EM algorithm would be:
\begin{description}
\item[E-Step]
\begin{equation}
p(X_{u,t} = x) = \frac{\lambda_{u,x} p(t|\theta_x)}{\sum_{x' \in \{g,c,s\}}\lambda_{u,x'} p(t|\theta_x')}
\end{equation}
\item[M-Step]
\begin{equation}
p(t|\theta_g) = 
\frac{\sum_{u \in G}c(t,u) p(X_{u,t} = g)}{\sum_{t' \in V}\sum_{u \in G}c(t',u) p(X_{u,t'} = g)}
\end{equation}
\begin{equation}
\lambda_{u,x} = 
\frac{\sum_{t \in V}c(t,u) p(X_{u,t} = x)}{\sum_{x' \in \{g,c,s\}}\sum_{t \in V}c(t,u) p(X_{u,t} = x')}
\end{equation}
\end{description}
After convergence of the EM algorithm, all the parameters are estimated including the group model, $\theta_g$, which is a distribution over terms representing shared group preferences as well as $\lambda_{u,g}$, $\lambda_{u,c}$, and $\lambda_{u,s}$ for each user $u$ which determine the contribution of each model in each user's preferences.

\smallskip 
In this section, we proposed a generative user model as a mixture of group, general (or collection), and specific models, and showed that the latent distribution of group model over terms can be extracted as the group's profile.

\section{Experiments}
\label{sec:exp}

In this section, we present our experiments to evaluate the effectiveness of the estimated language model of groups in the task of contextual suggestion. Furthermore, we analyse the effect of group granularity on group profiling. We first explain the data collection used in our experiments and then present the evaluation results. 

\subsection{Data Collection}
\label{sec:dataset}
\input{table_statistics}

In this research, we have made use of the TREC 2015  contextual suggestion\footnote{\url{https://sites.google.com/site/treccontext/trec-2015}} Batch task dataset. Contextual suggestion is the task of searching for complex information needs that are highly dependent on both context and user interests.  The dataset contains the information from 207 users including their age, gender, and set of rated places or activities as the user preferences (rates are in the range of -1 to 4). The task is to generate a list of ranked suggestions from a set of candidate attractions, by giving the user information as well as some information about the context, including location of trip, trip season, trip type, tripe duration, and the type of group the person is travelling with. For each user, we consider rated suggestions that are annotated with rates of more than 2 as relevant. Furthermore, we generate the user language models as a mixture of their relevant preferences considering the rates. Based on the information in the dataset, we divide users into several groups. Groupings are based on the users information and context information. Table~\ref{tbl:stat} presents grouping criteria, the groups, and number of users in each group.

\subsection{Group Profiling for Contextual Suggestion}
\label{sec:groupprofiling}

In this section, we investigate our second research question: ``\rqtwo''

We generate group-based rankings of suggestions to evaluate the quality of group profiles in content customization. 
To this end, one of the grouping approaches given in Table~\ref{tbl:stat} is chosen, e.g.\ based on users' age. Then we estimate language model of each group employing the approach explained in Section~\ref{sec:gp}. Afterward, regarding the information of the given request, i.e.\ the user information and context information, the group which the user belongs to is selected and based on the similarity of the language model of the selected group and the language model of candidate, the ranked list of the suggestions is generated.

Beside the group-based ranking, we generate a ranked list of suggestions based on the preferences of the user as a baseline. To do so, a language model is estimated as the mixture of the model of user preferences regarding their ratings and based on the similarity of the preferences language model and the candidate language model, a ranked list is generated.

Furthermore,  according to the explanation in Section~\ref{sec:gp}, the contribution of each of \emph{specific}, \emph{group}, and \emph{general} models in each user model is learned as the model parameters, i.e.\ $\lambda_{u,s}$, $\lambda_{u,g}$, and $\lambda_{u,c}$. Having these parameters empowers us to efficiently combine the group-based model with the preferences-based model for content customization. To this end, we smooth the preferences-based model of user with both the group model and the general model using JM-smoothing employing the learn parameters. To evaluate the quality of the combination, we have done experiments considering different grouping criteria.

\input{plot_perf}
Figure~\ref{fig:Chart1} presents the performance of employing different grouping approaches for group-\:based suggestion as well as preferences\:-\:based suggestion. The combinations of preferences\:-\:based suggestion and group-\:based suggestion are also reported.

As can be seen, among the group-based strategies, suggestions based on the duration of the trip is the most effective strategy. Also age of the user and the type of the group the user travels with, are rather important while type of the trip is not so important. This could be due to the fact that most of the time, the user's interests and beloved attractions do not change based on the type of trip which could be ``business'' or ``holiday''. 
On the other hand, combining the preferences-based suggestions with group-based suggestions in all grouping strategies leads to improvement. This means in case of incompleteness of user's profile, customizing the content based on the groups that user belongs to, implicitly fills the missing information and improves the performance of suggestions. However, this depends on the quality of the groups profiles that should reflect essential common (not general, not specific) characteristics of the groups.

\subsection{Effect of Group Granularity} 
\label{sec:gg}
In this section, we investigate our third research question: ``\rqthree''

In the grouping stage, sometimes users can be grouped based on different levels of granularity. For example, having the age of users, discretization can be done using binning with different sizes of bin. In this section, we analyse the effect of granularity of groups, and consecutively the size of the groups with a fixed volume of train data, on the quality of group profiling.

We have selected ``age'' of users as the grouping criterion and tried different bin sizes for discretization: 5 years, 10 years, 20 years and 40 years. 
Figure~\ref{fig:Chart2} shows the quality of groups profiles on different levels of granularity and consequently on different sizes of groups in the task of contextual suggestion. 
Each point in the figure represents a group of users and its position determines its size and the performance of group-based contextual suggestion for the users within the group. Moreover the horizontal lines represent overall performance of different levels of granularity. 
As can be seen, since the number of sample users is limited fine-grained grouping leads to having smaller groups. So small number of samples affects the group profiling quality and slightly decreases the performance. While coarse-grained grouping leads to having large groups that leads to not being able to adequately customize the group profile. 

In our dataset, 10 years granularity for ``age'' has the best performance since the formed groups are big enough so that the group profiling approach is able to estimate high quality models, and they are small enough so that the group profiles are easily distinguishable which leads to a more effective customization.

\input{plot_groupsize}

\section{Conclusions}
\label{sec:con}
%
In this paper, we dealt with the problem of group profiling. The main aim of this paper was \textsl{\maingoal}\

Our first research question was: \textsl{\rqone}\ 
We proposed to consider each user preferences as a mixture of general, specific and its group preferences and estimated the latent group preferences as the shared concerns among all group members. 

Our second research question was: \textsl{\rqtwo}\ 
We utilized the proposed group profiling approach for the task of contextual suggestion, and our experimental results showed that considering group-\:based suggestions along with user preferences\:-\:based suggestions can improve the content customization. 

Our third research question was: \textsl{\rqthree}\
We designed an experiment to investigate how group granularity may affect the group profiling quality. We found that the grouping approach should result in groups that are big enough to enable group profiling to infer high quality models and small enough to enable the extracted model to make customization for group members.


As the future work, we are going to find a way for learning how to employ group-based suggestions on the basis of different grouping criteria simultaneously. For example, how to combine suggestions based on the age with suggestions based on the gender. A further development would be to evaluate the proposed group profiling approach on other tasks and other kinds of data including non-textual data.


\mypar{Acknowledgments}  \small
This research is funded in part by the European Community's FP7 (project meSch, grant \# 600851) and the Netherlands Organization for Scientific Research (WebART project, NWO CATCH \# 640.005.001; ExPoSe project, NWO CI \# 314.99.108; DiLiPaD project, NWO Digging into Data \# 600.006.014).
%

\vspace{-10pt}

%
\renewcommand{\bibsection}{\section*{References}}
\bibliographystyle{abbrvnat}
\renewcommand{\bibfont}{\small\raggedright}
\setlength{\bibhang}{1em}
\setlength{\bibsep}{0pt} 
\bibliography{ref}  
%
%
\balancecolumns
\end{document}

%% file: table_statistics.tex
\newcommand*\rot{\rotatebox{90}}

\begin{table*}[!t]
\caption{\strut\label{tbl:stat} Statistics of users groups resulted by grouping based on different critera\strut}
\vspace{-10pt}
\small
\begin{tabularx}{\linewidth}{X|c@{~~}c@{~~}c@{~~}c@{~~}c|c@{~~}c|c@{~~}c@{~~}c@{~~}c|c@{~~}c@{~~}c|c@{~~}c@{~~}c@{~~}c|c@{~~}c@{~~}c@{~~}c}
\toprule
\bf Grouping Criterion 
& \multicolumn{5}{c|}{\textbf{Age}} 
& \multicolumn{2}{c|}{\textbf{Gender}} 
& \multicolumn{4}{c|}{\textbf{Group Type}} 
& \multicolumn{3}{c|}{\textbf{Trip Type}} 
& \multicolumn{4}{c|}{\textbf{Trip Duration}} 
& \multicolumn{4}{c}{\textbf{Season}}
\\
\midrule
Groups &
\rot{$<20$} & \rot{$20-30$} & \rot{$30-40$} & \rot{$40-50$} & \rot{$>50$} & \rot{male} & \rot{female} & \rot{alone} & \rot{family} & \rot{friends} & \rot{other} & \rot{holiday} & \rot{business} & \rot{other} & \rot{night out} & \rot{day trip} & \rot{weekend} & \rot{longer} & \rot{spring} & \rot{summer} & \rot{autumn} & \rot{winter}
\\
\midrule
Group Size (\#users) &
 9 & 87 & 60 & 22 & 23 & 107 & 91 & 23 & 101 & 65 & 4 & 176 & 7 & 12 & 8 & 17 & 99 & 74 & 20 & 99 & 52 & 22
\\
\bottomrule
\end{tabularx}
\vspace{-10pt}
\end{table*}

%% file: plot_perf.tex
    \definecolor{b}{HTML}{4981CE}
    \definecolor{g}{HTML}{859C27}
    \definecolor{r}{HTML}{B22222}
    \definecolor{o}{HTML}{FF6600}

\begin{figure}[!t]
\centering
\pgfplotsset{
    bar group size/.style 2 args={
        /pgf/bar shift={%
                -0.5*(#2*\pgfplotbarwidth + (#2-1)*\pgfkeysvalueof{/pgfplots/bar group skip})  + 
                (.5+#1)*\pgfplotbarwidth + #1*\pgfkeysvalueof{/pgfplots/bar group skip}},%
    },
    bar group skip/.initial=2pt,
    plot 0/.style={fill=b, draw=b, pattern color = b, pattern = north east lines},%
    plot 1/.style={fill=g, draw=g, pattern color = g, pattern = north west lines},%
    plot 2/.style={fill=r, draw=r, pattern color = r, pattern = dots},%
}
\begin{tikzpicture}
    \begin{axis}[
       width= 8.6cm, 
        height=7cm, 
        ybar,
        bar width= 7pt,
        enlarge y limits=0.2,
        ymajorgrids,
        minor tick num=1,
        x tick label style={rotate=45,anchor=east},
    xtick=data,
    anchor=north,
    ylabel={MAP},
  legend style={at={(0.965,-0.28)}
  ,legend columns=-1},
    nodes near coords,
    every node near coord/.append style={font=\tiny, rotate=90, anchor=west},
            xtick={1,2,3,4,5,6,7},
            xticklabels={
            age,
            gender,
            group,
            t-type,
            t-duration,
            season,
            preferences
        },
        ]
                
\addplot[plot 0,bar group size={0}{2}]
        coordinates 
        {
(1,0.3123)
(2,0.2417)
(3,0.2989)
(4,0.2122)
(5,0.3971)
(6,0.2879)
};
        \addplot[plot 1,bar group size={1}{2}]
        coordinates {
(1,0.6713)
(2,0.5187)
(3,0.5811)
(4,0.5066)
(5,0.5799)
(6,0.5562)
};

        \addplot[plot 2,bar group size={0}{1}]
        coordinates {
(7,0.5136)
};

\legend{Group, Group+Preferences, Preferences}

\end{axis}
\end{tikzpicture}
\vspace{-4pt}
\caption{
Performance of employing user preferences\:-\:based and group-\:based customization on contextual suggestion task. Improvements of combining group-\:based and preferences\:-\:based approach over the preferences\:-\:based approach and corresponding group-\:based approaches are statistically significant based on one-tailed t-test, with p-value $<$ 0.05. \label{fig:Chart1}}
 \vspace{-5pt}
 \end{figure}
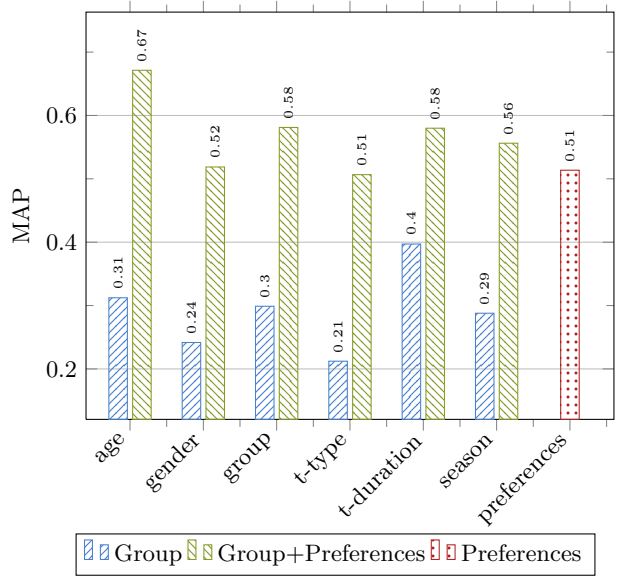

%% file: plot_groupsize.tex
\begin{figure}[!t]
\centering
\begin{tikzpicture}
\begin{axis}[
        width= 8.6cm, 
        height=6cm, 
        xmin=0,xmax=200,
        grid = both,
        ylabel={MAP},
        xlabel={Group Size},
        scatter/classes={
        	a={mark=square*,b},%
        	b={mark=triangle*,r},%
        	c={mark=o,draw=g},
        	d={mark=pentagon*,draw=o}
    	},
    	legend columns=2,
    	grid = both,
    	minor tick num=1,
	]

	\addplot[scatter,only marks,
		scatter src=explicit symbolic]
		coordinates {
			(9,0.2012)  [a]
			(44,0.2914) [a]
			(43,0.4181) [a]
			(45,0.2545)  [a]
			(15,0.2886)  [a]
			(14,0.2245)  [a]
			(8,0.1811)   [a]
			(23,0.3214) [a]
			(9,0.2012) [b]
			(87,0.2739) [b]
			(60,0.3855) [b]
			(22,0.3008) [b]
			(23,0.3214) [b]
			(96,0.2501) [c]
			(82,0.3300) [c]
			(23,0.3214) [c]
			(156,0.2418) [d]
			(46, 0.3633) [d]
		};
		\addplot[b,sharp plot,update limits=false] coordinates {(0,0.3003) (200,0.3003)} node[b!50!black, above] at (axis cs:165,0.3003) {\tiny{\textbf{0.3003}}}; 
		\addplot[r,sharp plot,update limits=false] coordinates {(0,0.3123) (200,0.3123)}node[r!50!black, above] at (axis cs:185,0.3123) {\tiny{\textbf{0.3123}}}; 
	    \addplot[g,sharp plot,update limits=false] coordinates {(0,0.2908) (200,0.2908)}node[g!50!black, above] at (axis cs:145,0.2908) {\tiny{\textbf{0.2908}}};  
		\addplot[o,sharp plot,update limits=false] coordinates {(0,0.2708) (200,0.2708)}node[o!50!black, above] at (axis cs:125,0.2708) {\tiny{\textbf{0.2708}}}; 
	
		\legend{5-years,10-years,20-years,40-years}
\end{axis}
\end{tikzpicture}
\vspace{-15pt}
\caption{
Effect of groups granularity on the performance group profiling
 \label{fig:Chart2}}
 \vspace{-15pt}
 \end{figure}
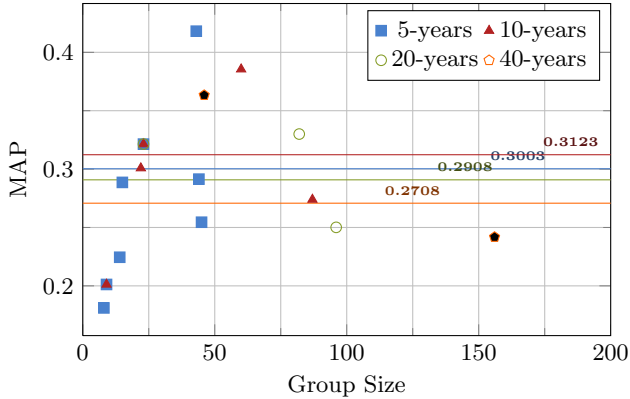